Snowmass White Paper

# Secondary Emission Calorimetry: Fast and Radiation-Hard

Fairfield University[1] - University of Iowa[2] - Argonne[3] Collaboration

A.Albayrak-Yetkin[2], B.Bilki[2,3], J.Corso[2], P.Debbins[2], G.Jennings[1], V.Khristenko[2], A.Mestvirisvilli[2], Y.Onel[2], I.Schmidt[2], C.Sanzeni[1], D.Southwick[2], D.R.Winn[1]*, T.Yetkin[2]

*Contact/PI - winn@fairfield.edu
cell/txt/vm 1+203.984.3993

*Abstract:* A novel calorimeter sensor for electron, photon and hadron energy measurement based on Secondary Emission(SE) to measure ionization is described, using sheet-dynodes directly as the active detection medium; the shower particles in an SE calorimeter cause direct secondary emission from dynode arrays comprising the sampling or absorbing medium. Data is presented on prototype tests and Monte Carlo simulations. This sensor can be made radiation hard at GigaRad levels, is easily transversely segmentable at the mm scale, and in a calorimeter has energy signal rise-times and integration comparable to or better than plastic scintillation/PMT calorimeters. Applications are mainly in the energy and intensity frontiers.

*Introduction:* Secondary Emission[1] (SE) is extraordinarily radiation resistant, and a fast process occurring in less than 1 ns in metal oxides. The SE beam monitors used at accelerators have no change in operation up to $10^{21}$ protons/cm$^2$ and use a simple native metal oxide film a few nm thick (alumina or titania)[2]. In PMT dynodes, the submicron thick metal-oxide coatings on the dynodes survive 10's GRad of heavily ionizing electron bombardment with tolerable degradation[3]. Essentially an SE calorimeter is composed of a monotonous array of PMT dynodes without a photocathode, which can self-operate at the Gain-BW of PMT, (~$10^6$ x 100 MHz). Dynode arrays can cycle to air without degradation, such as those used in mass-spectrometry detectors[4], and do not require thin film control as in photocathodes, making construction far easier and cheaper than PMT. Detectors in the forward regions at present and future colliders would greatly benefit from this device. Precision calorimetry at very high rates for lepton number violation would also benefit.

In a Secondary Emission (SE) calorimeter ionization detector module, Secondary Emission electrons (SEe) are generated from an SE surface/cathode/"dynodes", when charged hadronic or electromagnetic particles (shower particles) penetrate an SE sampling module either placed between absorber materials (Fe, Cu, Pb, W etc) in calorimeters, or as a homogeneous calorimeter consisting entirely of dynode sheets as the absorbers. An SE cathode – as on PMT dynodes - is a thin (10-50 nm thick) film. These films are typically simple metal-oxides $Al_2O_3$, MgO, CuO/BeO, or other higher yield materials. The simple as-found native oxide on Al produces 6 SEe/per incident at the peak of the SE yield parametrization as shown in Figs 2 below.

On the inner surface of a metal plate in vacuum, which serves as the entrance "window" to a compact vacuum vessel (metal or metal-ceramic), an SE film cathode is analogous to a photocathode, and the shower particles are similar to photons incident. The SEe produced from the top SE surface by the passage of shower particles, as well as the SEe produced from the passage of the shower particles through the dynodes, are similar to p.e. The SEe are then amplified by sheets of dynodes – metal-meshes or other planar dynodes (see Figs 2c, 3 below). The SEe yield δ is a strong function of momentum, following dE/dx as in the Sternglass[5] formulae. The yield follows a universal curve when normalized to the peak yield, as shown in Fig.1. As shown in Fig. 2 from CERN data, yields from a MIP on robust materials like alumina or titania films are only 1.10-1.2, requiring many dynodes for a MIP signal. On the other hand, as the shower is fully absorbed, those yields rise to about 6-7 at low energies. This variation with particle energy gives rise to quasi-compensation effects as the low energy nuclear fragments of hadron showers have high yields; for example 1 MeV alpha particle produces ~20 SEe.

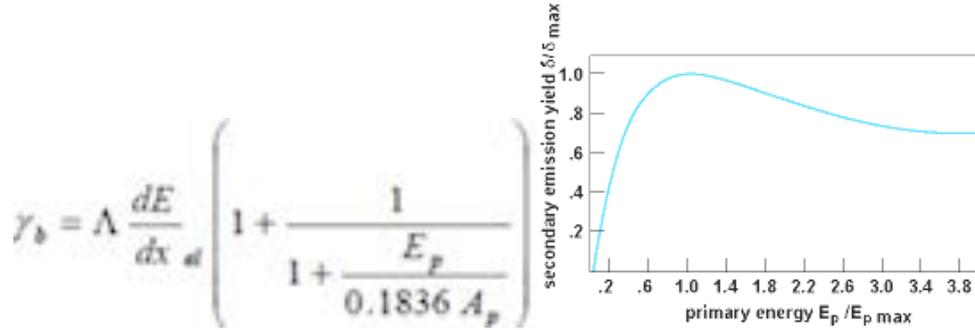

**Figs. 1:** SE Yield $\gamma_b$ from the Semiempirical Sternglass formula ($\Lambda$ = Escape Depth, $E_p$, $A_p$ = particle energy, mass) together with the universal curve of SE emission from metal oxides: the yield normalized by the maximum yield vs the particle energy normalized to the voltage/energy of the maximum yield emission.

$$\gamma_b = \Lambda \frac{dE}{dx}\bigg|_{el} \left[ 1 + \frac{1}{1 + \frac{E_p}{0.1836 A_p}} \right]$$

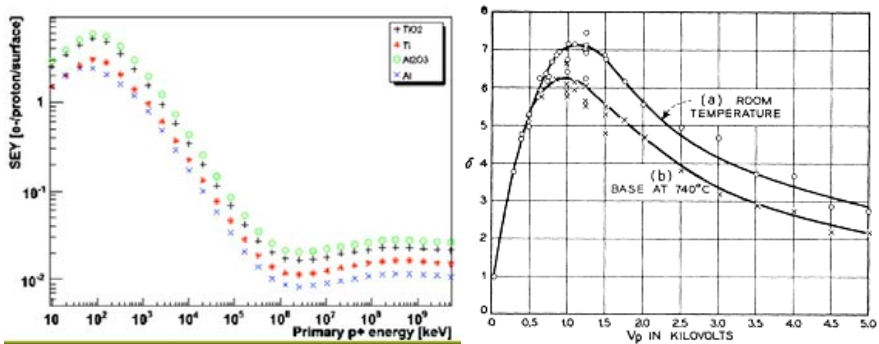

**Fig 2a,b:** Yield vs proton energy for titania and alumina and for the bare metals; Yield for MgO vs electron energy.

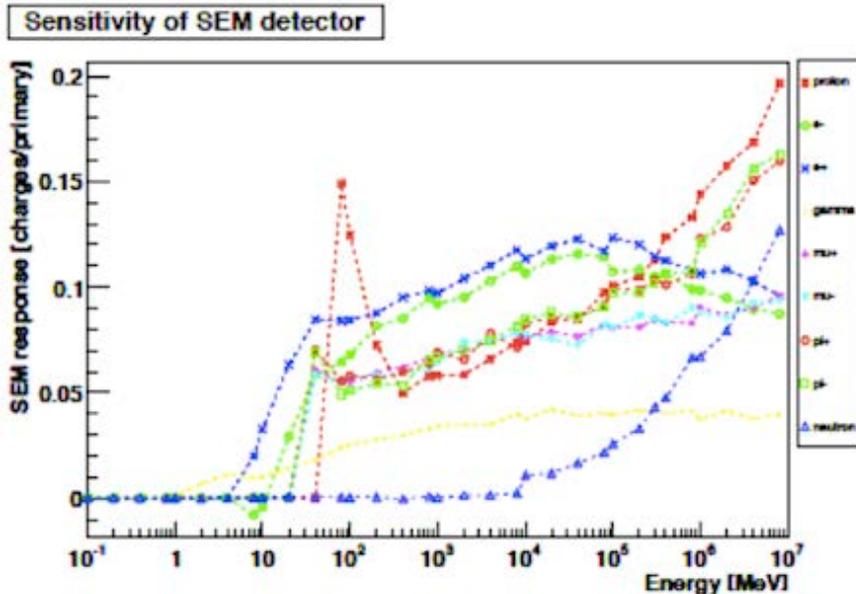

**Fig. 2c:** The number of SE electrons per primary collected in the CERN BLMS system test cell (10 nm thick alumina SE surface) vs primary energy- p (red square), e-, e+( green-solid, blue-x), μ+, μ-(purple, magenta), pi+, pi- (red-circle, green-open box), n(blue triangle), gamma(yellow) primaries. The higher yields above ~10 GeV are speculated to be from delta-rays. From Ref[2] B. Dehning et al. ACHTUNG! The energy bins < ~30 MeV have a threshold from the thick metal entrance window – the regions below 1 MeV have the highest yields; the negative signal at ~8 MeV is absorption of e- inside the signal electrode.

SE sensor modules can make use of electrochemically etched/machined or laser-cut metal mesh dynode sheets, as large as ~30 cm square or more, to amplify the Secondary Emission Electrons (SEe), much like those that compact metal mesh or mesh dynode PMT's use to amplify p.e.'s. The secondary emission yield follows dE/dx vs E or p. MIPs typically saturate at 0.15 SEe/mip, and rise by a factor of 20 at the low energy peak, typically 4-7 SE/incident particle, although thin film synthetic diamond can emit up to 120 SEe at 3 KV incident. We have studied this using GEANT4[6] to MC as below.

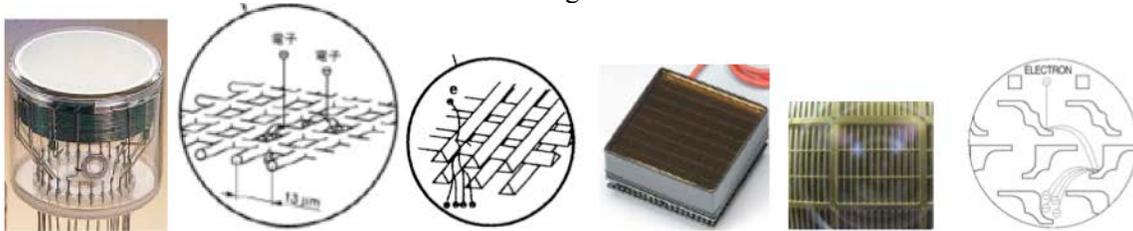

*Figures 3a:* *Left to right top* – mesh dynode PMT with typical meshes – note the 11 μm scale of the mesh – and PMT using etched metal dynodes, as formed in sheets.

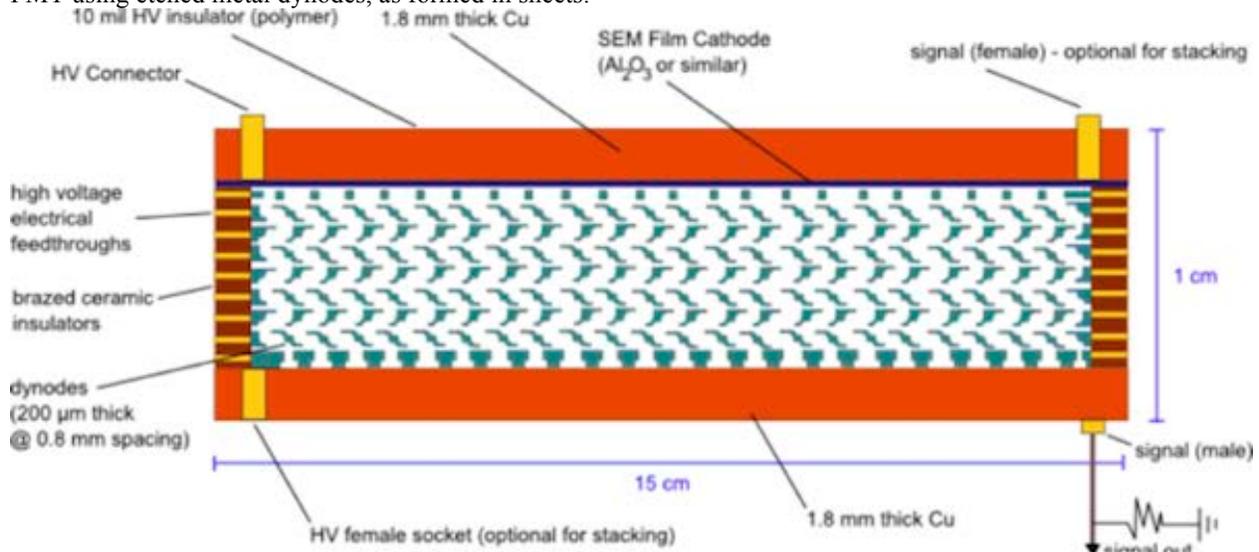

*Figure 3b:* A cartoon of a Secondary Emission calorimeter module, using etched metal sheet dynodes similar to those used by Hamamatsu. Commercial Cu-Be meshes are also possible and low cost ($20/m$^2$ for appropriate fine metal woven sheets).

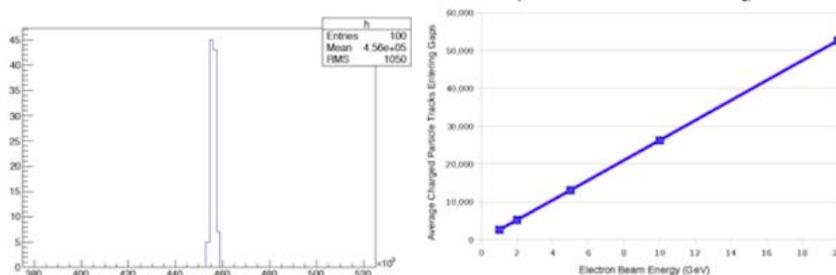

*Figures 4:* (Left) GEANT4-predicted histogram of the number of secondary emitted electrons by shower particles in an e-m showers initiated by 100 GeV electrons (100 events), in a *homogeneous mesh (no absorber plates) calorimeter* consisting of 10μm wire W-mesh, 30% transparent, spaced at 10 μm calorimeter using the Sternglass formula for SEe yields, and assuming an SE yield for 100 eV electrons of

3. The stochastic term inferred is less than $1\%/\sqrt{E(GeV)}$ – (no systematics nor various noise issues are in the MC to date). The linearity (right) is outstanding. The density is about 40% of pure W.

The construction requirements for an SE Sensor Module are much easier than a PMT, since:
1. the entire final assembly can be done in air;
2. there are no critical controlled thin film vacuum depositions for a photocathode, cesiation or other oxygen-excluded processes or other required vacuum activation is not necessary (although possibly desired for enhanced performance).
3. bake-out can be a refractory temperatures, unlike a photocathode which degrades at T>300°C;
4. the SE module is sealed by normal vacuum techniques (welding, brazing, diffusion-bonding or other high temperature joinings), with a simple final heated vacuum pump-out and tip-off.

The modules envisioned are compact, high gain, high speed, exceptionally radiation damage resistant, rugged, and cost effective, and can be fabricated in arbitrary tileable shapes. Mesh dynodes will work at 10% gain in 1.2 T at 45° to the B-field[7]. The SE sensor module anodes can be segmented transversely to sizes appropriate to reconstruct electromagnetic shower cores with high precision. The GEANT4 and existing calorimeter data estimated calorimeter response performance is between 35-50 Secondary Emission electrons (SEe) per GeV of electromagnetic deposition, in a calorimeter with 1 cm thick Cu absorbers interspersed with 15 planes of Cu mesh dynodes, with a gain per SEe of $\sim 10^5$ per SEe, and an e/pi<1.2. The calorimeter pulse width 10%-10% (90%) is estimated to be <11 ns based on shower length, secondary electron transit time estimate, and mesh PMT TTS. A MC of a *homogeneous* (i.e. no absorber except dynode sheets) calorimeter using 10μm W mesh sheets at 10 μm spacing (about 40% of the density of W) generates in excess of 40,000 SEe per GeV, and a stochastic term inferred of less than $1\%/\sqrt{E(GeV)}$, assuming each SEe is amplified by the downstream mesh to a conveniently detectable level ($g > 4 \times 10^4$ per SEe). We emphasize that an SEe is treated exactly like a p.e. in a scintillator calorimeter. In a scintillator calorimeter, many photons are created, but typically 0.1-1% are collected and converted to p.e. by a PMT of SiPM; by contrast, in an SE calorimeter, relatively few SE electrons are created, but almost all are collected and amplified by the dynode stacks as SEe.

*Initial Tests of SE Calorimetry:* We have tested this concept at a CERN test beam in 2011 - using metal mesh PMT as shown in Fig 4 below. The photocathode was disabled by using a +HV base, operating the anode at ~ +2KV, D1 at ground, and the photocathode at small positive voltages. Blue LED showed that the photocathode was insensitized. The tube was operated with particles entering from the base side, with the photocathode window blackened. The tube was exposed to beams of 225 GeV muons and 100 GeV electrons entering a 3cm Pb radiator, to simulate shower max.

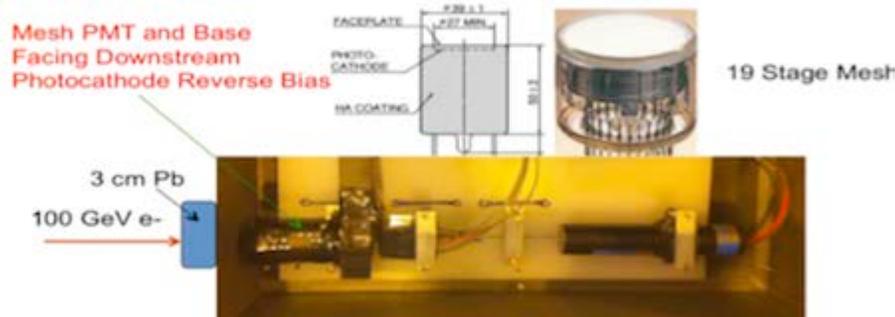

*Fig 5: Beam Test Configuration for SE emission signals:* Hamamatsu 19 stage mesh PMT in the CERN test beamline, in the phototube test box. 225 GeV μ+ or 100 GeV e- hitting 3 cm of Pb radiator incident from left. The photocathode was disabled using a +HV base, anode at +2.1KV, Gain=8x10^4, D1= ground; Photocathode = +10V through 400kΩ to ground.

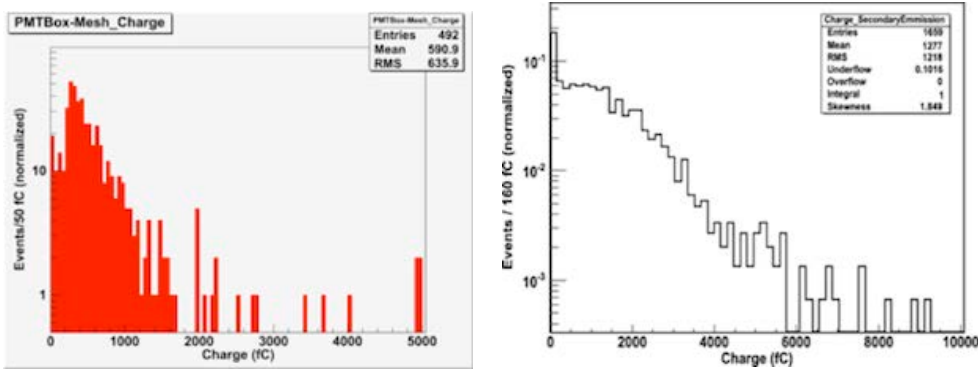

*Fig 6(L):* **Beam test results of Mesh SEe "sensor"**: 100 GeV electrons into 3 cm Pb, with one SE sampling "cell". Peak in mesh PMT signal corresponds to 41 SE electrons (mesh gain ~$10^5$) at about shower max. The large width is essentially entirely consistent with shower width fluctuations on a single shower sampling after 3 cm Pb. **Fig 6(R): Detection efficiency** - Muon Efficiency as selected by wire chambers; the peak at zero is below threshold indicating:~75% detection efficiency for a single particle through at most the first 5 meshes – after that, the gain falls below threshold. (Cu-Be Mesh PMT: 19 stages, g~2 x$10^6$; 1st 5 mesh ~ 80% effy at 20%/mesh). We infer that the yield on a MIP is 1.11 per SE surface in line with measurements from CERN as in Figs 2a, 2c.

The peak of the signal in the mesh PMT signal corresponds to 41 SE electrons (mesh gain ~$10^5$) at about shower max. The large width is essentially consistent with shower fluctuations on a single shower sampling after 3 cm Pb, over the 1.25" PMT face. This implies > 600 SEe/GeV is possible (i.e treated the same as ~600 pe/GeV in a scintillator calorimeter) with a full calorimeter.

The detected muon signal was 74% efficient, consistent with 3 pe, consistent with MIP delta rays/SE electrons>50 eV from the photocathode surface, and SE from the 19 dynodes. The electron shower max was fully efficient, with a mean number of SEe of ~30-40, within a factor of 2 expected for a sample of a 100 GeV shower at 5.5 Lrad and the dynode stack diameter. This implies with homogeneous mesh stack calorimeter (no other absorber) that >600 SEe per GeV are possible with the particular mesh used in the mesh PMT, consistent with the MC for this Cu/Be mesh, and a resolution of ~4%/√E.

Another test beam was run at CERN in 2012 using an array of metal mesh (19 stage) PMT with the Photocathode shut off, as assembled in groups of 9 into SE sensor modules. These modules are shown in the figure below. The SE sensor modules were interspersed with 1.75Xo Fe slabs as shown below. A variable length of Fe slabs (0-9Xo)were placed in front, and data taken with each thickness in front. The ADC histogram data with the MC ADC histogram of the set-up are shown below, and the MC well-matches the ADC response of the SE sensor modules.

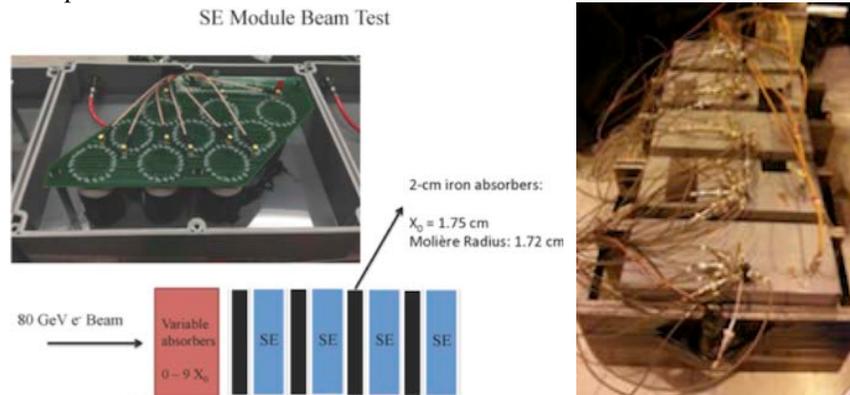

**Figures 7:** Left – an SE sensor module using 9 mesh dynode tubes with the photocathode shut off. Right: 5 sensor modules (as at left) stacked between 1.75Xo Fe absorbers. As shown above, data were taken with 80 GeV electrons incident, varying the absorber thickness ahead of the modules, in order to simulate a full calorimeter.

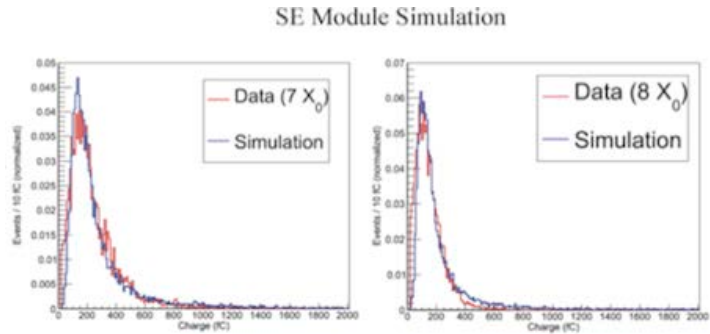

**Figures 8:** Data compared with MC of the ADC histogram response of the SE sensors to 80 GeV electrons for showers after 7Xo and after 8 Xo in front of the SEsensor stacks. Good agreement with the simulation.

We then used the MC to predict the energy resolution of a possible full calorimeter for electrons using this crude calorimeter (1.75 Xo, with only the partial areal coverage per module of the showers as shown by the module construction above) sampling, and assuming enough modules to cover 30 Xo. These MC predictions are shown below.

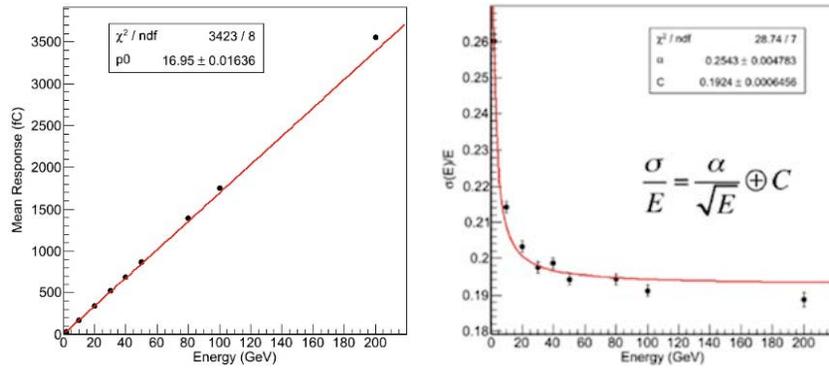

**Fig. 9:** Electron linearity and energy resolution in GEANT4 simulations of a 1.75 Xo FE sampling SEsensor calorimeter.

GEANT4 was tuned to simulation a "homogeneous" SE Calorimeter – a stack of W or Cu dynodes made of mesh, 10 µm in thickness spaced at 10 µm (about 40% of the density of W or Cu), with no other absorber. The predicted resolution is ~1-3%/√E, assuming each SEe is treated like a p.e.

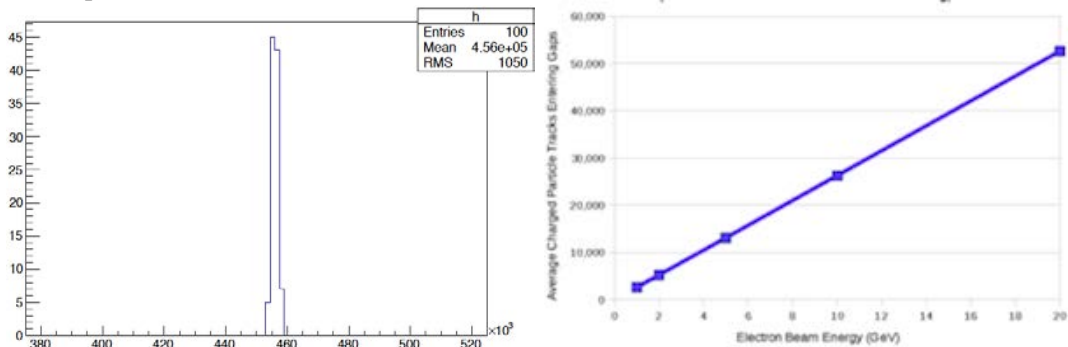

***Figs.10 (as 4 above):*** Left: GEANT4-predicted histogram of the number of secondary emitted electrons created by shower particles ($e^-$ or $e^+$) in e-m showers of 100 GeV electrons (100 events), incident on a 10µm W-wire mesh spaced by 10 µm – a *homogeneous* – no absorber plates – calorimeter, using the Sternglass formula for SEe yields on an oxide film with a peak yield of 3 at 100 V. Right: GEANT4 predicted linearity on e-m showers for a homogeneous W-mesh calorimeter with 40% of the density of W.

*Future Directions:*
*Neutrons:* Neutrons are an especially interesting application since the dynodes arranged as a dense stack can be individually coated on the back-side with a 1-10 micron thick $Li^6B^{10}$ film. The film thickness would be adjusted to optimize alpha yield into vacuum from the absorption of neutrons in the LiB, a tradeoff between thin so that the alpha resulting from neutron absorption escapes to cause secondary emission in neighboring dynodes, and thick to absorb neutrons. Alpha particles of 0.1-5 MeV have very high secondary yield (~20-80) as the specific ionization in the oxide films is very high[8]. This has the potential for a unique neutron detector, and for HEP calorimetry where enhanced neutron detection may result in improved compensation.

*High Secondary Yield Materials:* We show classic high SE yield materials below[9] (Fig.11). Synthetic diamond films have extraordinary secondary yields[10], and are an obvious choice for higher performance, with a minimal number of dynode sheets(Fig 12).

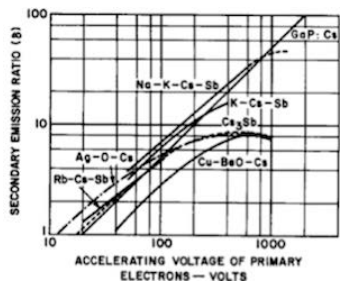

*Fig. 11:* Secondary yield of common materials[9].

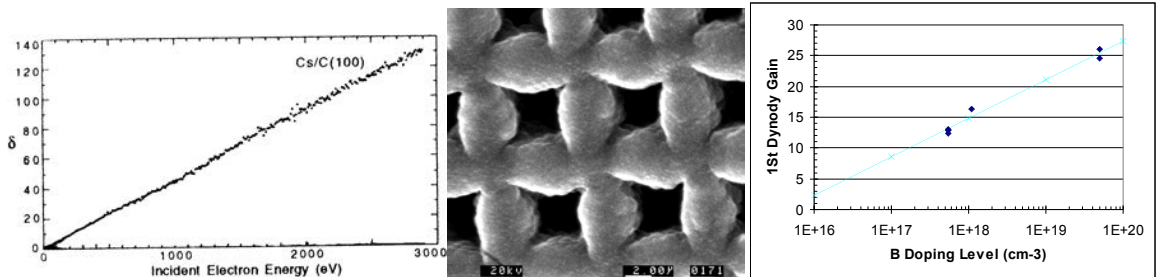

*Figs 12:* Synthetic Diamond SE yields: Left - Up to 130 at 3KV (data from Naval Research Labs); Middle - Mesh coated with 1 μm of diamond. Center-center=6 μm. Right – Diamond Dynode SE yield at 300 V vs B-doping level. ~25 per incident electron is easily available. (Data from NanoSciences, Oxford, CT and from Burle Electro-Optics)

**Summary:** The purpose of secondary emission calorimeters is particle (e, gamma, n, p, meson, nuclei) energy measurements in extreme radiation environments, and at very high rates. Benefits include fine segmentation, close packing and low cost. The modules envisioned are densely compact, high gain, high speed, exceptionally radiation damage resistant, rugged, and cost effective, and can be fabricated in arbitrary tileable shapes, to form large nearly hermetic detector arrays. The SE sensor module anodes can be segmented transversely to sizes appropriate to reconstruct incident gamma or neutron direction, with sufficient readout. Neutron detection may be enhanced by thin film coatings. Additional benefits outside of HEP include operation in harsh environments (space environs, high temperature/pressure environs e.g. boreholes>1 km deep, high g-forces), imaging detectors in medicine and industry, radiocontraband detection, and beyond.